# A New Approach to Population Sizing for Memetic Algorithms: A Case Study for the Multidimensional Assignment Problem*

**D. Karapetyan**                                Daniel.Karapetyan@gmail.com
Department of Computer Science, Royal Holloway University of London, Egham, Surrey, TW20 0EX, UK

**G. Gutin**                                            G.Gutin@cs.rhul.ac.uk
Department of Computer Science, Royal Holloway University of London, Egham, Surrey, TW20 0EX, UK

**Abstract**

Memetic Algorithms are known to be a powerful technique in solving hard optimization problems. To design a memetic algorithm one needs to make a host of decisions; selecting a population size is one of the most important among them. Most algorithms in the literature fix the population size to a certain constant value. This reduces the algorithm's quality since the optimal population size varies for different instances, local search procedures and running times. In this paper we propose an adjustable population size. It is calculated as a function of the running time of the whole algorithm and the average running time of the local search for the given instance. Note that in many applications the running time of a heuristic should be limited and therefore we use this limit as a parameter of the algorithm. The average running time of the local search procedure is obtained during the algorithm's run. Some coefficients which are independent with respect to the instance or the local search are to be tuned before the algorithm run; we provide a procedure to find these coefficients.

The proposed approach was used to develop a memetic algorithm for the Multidimensional Assignment Problem (MAP or $s$-AP in the case of $s$ dimensions) which is an extension of the well-known assignment problem. MAP is NP-hard and has a host of applications. We show that using adjustable population size makes the algorithm flexible to perform well for instances of very different sizes and types and for different running times and local searches. This allows us to select the most efficient local search for every instance type. The results of computational experiments for several instance families and sizes prove that the proposed algorithm performs efficiently for a wide range of the running times and clearly outperforms the state-of-the art 3-AP memetic algorithm being given the same time.

## 1 Introduction

A memetic algorithm is a combination of an evolutionary algorithm with a local search procedure (Krasnogor and Smith, 2005). The memetic approach is a template for an algorithm rather than a set of rules for designing a powerful heuristic. A typical frame of a memetic algorithm is presented in Figure 1 (for a formal definition of a memetic algorithm main loop see, e.g., Krasnogor and Smith (2008)).

---

*A preliminary version of this paper was accepted for publication in proceedings of the Stochastic Local Search Conference 2009 in Lecture Notes in Computer Science (Gutin and Karapetyan, 2009b).

1. Produce the first generation, i.e., a set of feasible solutions.
2. Apply a local search procedure to every solution in the first generation.
3. Repeat the following while a termination criterion is not met:
    (a) Produce a set of new solutions by applying so-called genetic operators to solutions from the previous generation.
    (b) Improve every solution in this set with the local search procedure.
    (c) Select several best solutions from this set to the next generation.

Figure 1: A typical memetic algorithm frame.

When implementing a memetic algorithm, one faces a lot of questions. Some of these questions, like selecting the most appropriate local search or crossover operators, were widely discussed in the literature while others are still not investigated enough. In this research we focus our attention on the population sizing.

Population size is the number of solutions (chromosomes) maintained at a time by a memetic algorithm. Many researchers indicate the importance of selecting proper population sizes (Glover and Kochenberger, 2003; Harik et al., 1999; Hart et al., 2005). However, the most usual way to define the population size is to fix it to some constant at the design time (Cotta, 2008; Grefenstette, 1986; Hart et al., 2005; Huang and Lim, 2006). Several more sophisticated models based on statistical analysis of the problem or self-adaptive techniques are proposed for genetic, particle swarm optimization and some other evolutionary algorithms (Cotta, 2008; Eiben et al., 2004; Goldberg et al., 1991; Harik et al., 1999; Hart et al., 2005; Kaveh and Shahrouzi, 2007; Lee and Takagi, 1993) but they all are not suitable for memetic algorithms because of the totally different algorithm dynamics.

It is known (Hart et al., 2005) that in memetic algorithms the population size, the solution quality and the running time are mutually dependent. Often the population size is fixed at the design time which, for a given algorithm with a certain termination criterion, determines the solution quality and the running time. However, in many applications it is the running time which has to be fixed. This leads to a problem of finding the most appropriate population size $m$ for a fixed running time $\tau$ such that the solution quality is optimized. However, the population size $m$ depends not only on the given time $\tau$ but also on the instance type and size, on the local search performance and on the computational platform. The fact that the optimal population size depends on the particular instance, forces researchers to use parameter control to adapt dynamically the population size for all the factors during the run (see, e.g., Coelho and de Oliveira (2008); Eiben et al. (2004); Kaveh and Shahrouzi (2007)). However, none of these approaches consider the running time of the whole algorithm and, hence, are poorly suitable for a strict time limitation.

Instead of it, we have found a parameter encapsulating all these factors, i.e, a parameter which reflects on the relation between the instance, the local search procedure

New Approach to Population Sizing: Case Study for MAP

and the computation platform. It is the average running time $t$ of the local search procedure applied to some solutions of the given instance. Definitely this time depends on the particular solutions but later we will show that $t$ can be measured at any point of the memetic algorithm run with a good enough precision.

Now we can find a near-optimal population size $m_{\text{opt}}$ as a function of $\tau$ and $t$. In particular, it can be calculated as

$$m_{\text{opt}}(\tau, t) = a \cdot \frac{\tau^b}{t^c},$$

where $a$, $b$ and $c$ are some tuned (Eiben et al., 1999) constants which reflect on the specifics of the other algorithm factors.

Observe that this is not a pure parameter tuning. Indeed, the population size depends on the average local search running time $t$ which is obtained during the algorithm run. Thus, our approach is a combination of the parameter tuning and control.

In our previous attempt to adjust the population size (Gutin and Karapetyan, 2010) we assumed that it depends on the instance size $n$ only (i.e., $m = m(n)$) but an obvious disadvantage of this approach is that it does not differentiate between instance types.

In this paper the proposed approach is applied to the Multidimensional Assignment Problem. We think that the obtained results can be extended to many hard optimization problems. The expression for $m_{\text{opt}}(\tau, t)$ above follows a natural rule that the population size should be increased if the algorithm is given more time and decreased if local search is slower. Even if this formula is not appropriate in some cases, we believe that the main idea of calculating the population size before the algorithm run as a function of the given time and the running time of local search should be suitable for virtually any problem.

The *Multidimensional Assignment Problem* (MAP) (abbreviated $s$-AP in the case of $s$ dimensions, also called *(axial) Multi Index Assignment Problem*, MIAP, (Bandelt et al., 2004; Pardalos and Pitsoulis, 2000a)) is a well-known optimization problem. It is an extension of the *Assignment Problem* (AP), which is exactly the two dimensional case of MAP. While AP can be solved in the polynomial time (Kuhn, 1955), $s$-AP for every $s \geq 3$ is NP-hard (Garey and Johnson, 1979) and inapproximable (Burkard et al., 1996b)[1]. The most studied case of MAP is the case of three dimensions (Aiex et al., 2005; Andrijich and Caccetta, 2001; Balas and Saltzman, 1991; Crama and Spieksma, 1992; Huang and Lim, 2006; Spieksma, 2000) though the problem has a host of applications for higher numbers of dimensions, e.g., in matching information from several sensors (data association problem), which arises in plane tracking (Murphey et al., 1998; Pardalos and Pitsoulis, 2000b), computer vision (Veenman et al., 2003) and some other applications (Andrijich and Caccetta, 2001; Bandelt et al., 2004; Burkard and Çela, 1999), in routing in meshes (Bandelt et al., 2004), tracking elementary particles (Pusztaszeri et al., 1996), solving systems of polynomial equations (Bekker et al., 2005), image recognition (Grundel et al., 2004), resource allocation (Grundel et al., 2004), etc.

For a fixed $s \geq 2$, the problem $s$-AP is stated as follows. Let $X_1 = X_2 = \ldots = X_s = \{1, 2, \ldots, n\}$; we will consider only vectors that belong to the Cartesian product $X = X_1 \times X_2 \times \ldots \times X_s$. Each vector $e \in X$ is assigned a non-negative weight $w(e)$. For a vector $e \in X$, the component $e_j$ denotes its $j$th coordinate, i.e., $e_j \in X_j$. A collection $A$ of $t \leq n$ vectors $A^1, A^2, \ldots, A^t$ is a *(feasible) partial assignment* if $A_j^i \neq A_j^k$ holds for each $i \neq k$ and $j \in \{1, 2, \ldots, s\}$. The *weight* of a partial assignment $A$ is $w(A) = \sum_{i=1}^{t} w(A^i)$. A partial assignment with $n$ vectors is called *assignment*. The objective of $s$-AP is to find an assignment of minimal weight.

---

[1] Burkard et al. show it for a special case of 3-AP and since 3-AP is a special case of $s$-AP the result can be extended to the general MAP.

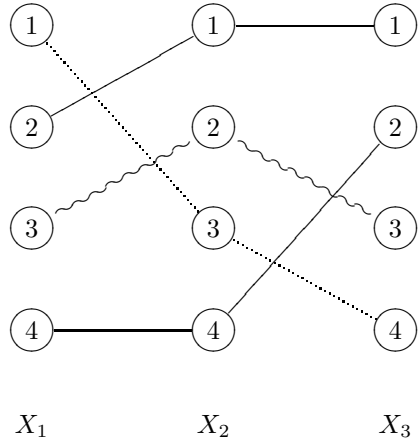

Figure 2: An example of an assignment for a problem with $s = 3$ and $n = 4$. This assignment contains the following vectors: (1, 3, 4), (2, 1, 1), (3, 2, 3) and (4, 4, 2).

A graph formulation of the problem (see Fig. 2) is as follows. Having an $s$-partite graph $G$ with parts $X_1, X_2, \ldots, X_s$, where $|X_i| = n$, find a set of $n$ disjoint cliques in $G$ of the minimal total weight if every clique $Q$ in $G$ is assigned a weight $w(Q)$ (note that in the general case $w(Q)$ is not simply a function of the edges of $Q$).

An integer programming formulation of the problem can be found in (Gutin and Karapetyan, 2009a).

Finally we provide a *permutation form* of the assignment which is sometimes very convenient. Let $\pi_1, \pi_2, \ldots, \pi_s$ be permutations of $X_1, X_2, \ldots, X_s$, respectively. Then $\pi_1 \pi_2 \ldots \pi_s$ is an assignment of weight $\sum_{i=1}^{n} w(\pi_1(i)\pi_2(i) \ldots \pi_s(i))$. It is obvious that some permutation, say the first one, may be fixed without any loss of generality: $\pi_1 = 1_n$, where $1_n$ is the identity permutation of $n$ elements. Then the objective of the problem is as follows:

$$\min_{\pi_2, \ldots, \pi_s} \sum_{i=1}^{n} w(i \pi_2(i) \ldots \pi_s(i))$$

and it becomes clear that there exist $n!^{s-1}$ feasible assignments and the fastest known algorithm to find an optimal assignment takes $O(n!^{s-2} n^3)$ operations. Indeed, without loss of generality set $\pi_1 = 1_n$ and for every combination of $\pi_2, \pi_3, \ldots, \pi_{s-1}$ find the optimal $\pi_s$ by solving corresponding AP in $O(n^3)$.

Thereby, MAP is very hard; it has $n^s$ values in the weight matrix, there are $n!^{s-1}$ feasible assignments and the best known algorithm takes $O(n!^{s-2} n^3)$ operations. Compare it, e.g., with the Travelling Salesman Problem which has only $n^2$ weights, $(n-1)!$ possible tours and which can be solved in $O(n^2 \cdot 2^n)$ time (Held and Karp, 1962).

The problem described above is called *balanced* (Clemons et al., 2004). Sometimes MAP is formulated in a more general way if $|X_1| = n_1$, $|X_2| = n_2$, ..., $|X_s| = n_s$ and the requirement $n_1 = n_2 = \ldots = n_s$ is omitted. However this case can be easily transformed into the balanced problem by computing $n = \max_i n_i$ and complementing the weight matrix to an $n \times n \times \ldots \times n$ matrix with zeros.

MAP was studied by many researchers. Several special cases of the problem were intensively studied in the literature (see Kuroki and Matsui (2007) and references there) but only for a few classes of them polynomial time exact algorithms were found, see, e.g., Burkard et al. (1996a,b); Isler et al. (2005). In many cases MAP remains hard to solve (Burkard et al., 1996b; Crama and Spieksma, 1992; Kuroki and Matsui, 2007; Spieksma and Woeginger, 1996). For example, if there are three sets of points of size $n$ on a Euclidean plain and the objective is to find $n$ triples, every triple has a point in each set, such that the total circumference or area of the corresponding triangles is minimal, the corresponding 3-APs are still NP-hard (Spieksma and Woeginger, 1996). Apart from proving NP-hardness, researches studied asymptotic properties of some special instance families (Grundel et al., 2004; Gutin and Karapetyan, 2009c).

As regards the solution methods, there exist exact and approximation algorithms (Balas and Saltzman, 1991; Crama and Spieksma, 1992; Kuroki and Matsui, 2007; Pasiliao et al., 2005; Pierskalla, 1968) and heuristics including construction heuristics (Balas and Saltzman, 1991; Gutin et al., 2008; Karapetyan et al., 2009; Oliveira and Pardalos, 2004), greedy randomized adaptive search procedures (Aiex et al., 2005; Murphey et al., 1998; Oliveira and Pardalos, 2004; Robertson, 2001) (including several concurrent implementations, see, e.g., Aiex et al. (2005); Oliveira and Pardalos (2004)) and a host of local search procedures (Aiex et al., 2005; Balas and Saltzman, 1991; Bandelt et al., 2004; Burkard et al., 1996b; Clemons et al., 2004; Gutin and Karapetyan, 2009a; Huang and Lim, 2006; Oliveira and Pardalos, 2004; Robertson, 2001).

Construction heuristics give us flexibility to generate a solution with some certain quality requirements (in case of approximation algorithms one can even get some quality guarantee). Using a local search algorithm, one is able to further improve the solution. However, a standard local search can optimize the solution only to a local minimum and no further improvements are available after that. Variable neighborhood search (see, e.g., Talbi (2009)) yields more powerful algorithms (Gutin and Karapetyan, 2009a) which, though, still have properties of local search. In order to improve the solution even more one should use more powerful metaheuristics.

Two metaheuristics were proposed for MAP in the literature, namely a simulated annealing procedure (Clemons et al., 2004) and a memetic algorithm (Huang and Lim, 2006). The purpose of this research is to develop a new approach in designing memetic algorithms and to test it in the case of MAP. We show that our approach improves existing results and the obtained heuristic is suitable for relatively large instances. It is flexible in choosing 'solution quality'/'running time' balance as well as in selecting the most appropriate local search for every instance type.

The rest of the paper is organized as follows. The proposed approach to the population sizing is described Section 2. The details of the memetic algorithm designed for MAP are discussed in Section 3. The test bed for our computational experiments is introduced in Section 4. The experiment results are provided and discussed in Section 5. Apart from the designed memetic algorithm, we evaluate two other MAP metaheuristics known from the literature and compare the results. The main outcomes of the presented research are summarized in Section 6.

## 2 Managing Solution Quality and Population Sizing

Having some fixed procedures for production of the first generation (Step 1 in Figure 1), improving a solution (Steps 2 and 3b) and obtaining the next generation from the previ-ous one (Steps 3a and 3c), the algorithm designer is able to manage the solution quality and the running time of the algorithm by varying the termination criterion (Step 3) and the population size, i.e., the number of maintained solutions in Steps 1 and 3c.

Usually, a termination condition in a memetic algorithm tries to predict the point after which any further effort is useless or, at least, not efficient. A typical approach is to count the number $I_{\text{idle}}$ of running generations which did not improve the best result and to stop the algorithm when this number reaches some predefined value. A slightly more advanced prediction method is applied in the state-of-the-art algorithm for the Generalized Traveling Salesman Problem by Gutin and Karapetyan (2010). It stops the algorithm when $I_{\text{idle}}$ reaches $k \cdot I_{\text{prev}}$, where $k > 1$ is a constant and $I_{\text{prev}}$ is the maximum $I_{\text{idle}}$ obtained before the current solution was found.

In case of such termination conditions, the running time of the algorithm is unpredictable and, hence, cannot be adjusted for one's needs. Observe that many applications (like real-time systems) in fact have strict time limitations. To satisfy these limitations, we bound our algorithm within some fixed running time and aim to use this time with the most possible efficiency. Below we discuss how the parameters of the algorithm should be adjusted for this purpose.

## 2.1 Population Size

Population size is the number of solutions maintained by a memetic algorithm at the same time. This number may vary from generation to generation but we decided to keep the population size constant during the algorithm run in order to simplify the research.

Let $I$ be the total number of generations during the algorithm run and $m$ be the population size. Then the running time of the whole algorithm is proportional to $I \cdot m$. Indeed, the most time consuming part of a memetic algorithm is local search. The number of times the local search procedure is applied is proportional to $I \cdot m$, and we have shown empirically (see Figure 3) that the average running time of a local search depends only marginally on the population size. Since we fix the running time of the whole algorithm, we get:

$$I \cdot m \approx \text{const}.$$

In other words, we claim that inversely proportional change of $I$ and $m$ preserves the running time of the whole algorithm; our experiments confirm it.

Since $I \cdot m = \text{const}$, we need to find the optimal ratio between $I$ and $m$. Our experimental analysis shows that this ratio is crucial for the algorithm performance: for a wrongly selected ratio between $I$ and $m$, the relative solution error, i.e., the percentage above the optimal objective value, may be twice as big as the relative solution error for a well fitted ratio, see Figure 4.

Observe that the optimal ratio between $I$ and $m$ depends on the following factors:

- Given time $\tau$;

- Instance type and size;

- Computational platform;

- Local search procedure;

- Genetic operators and selection strategies.

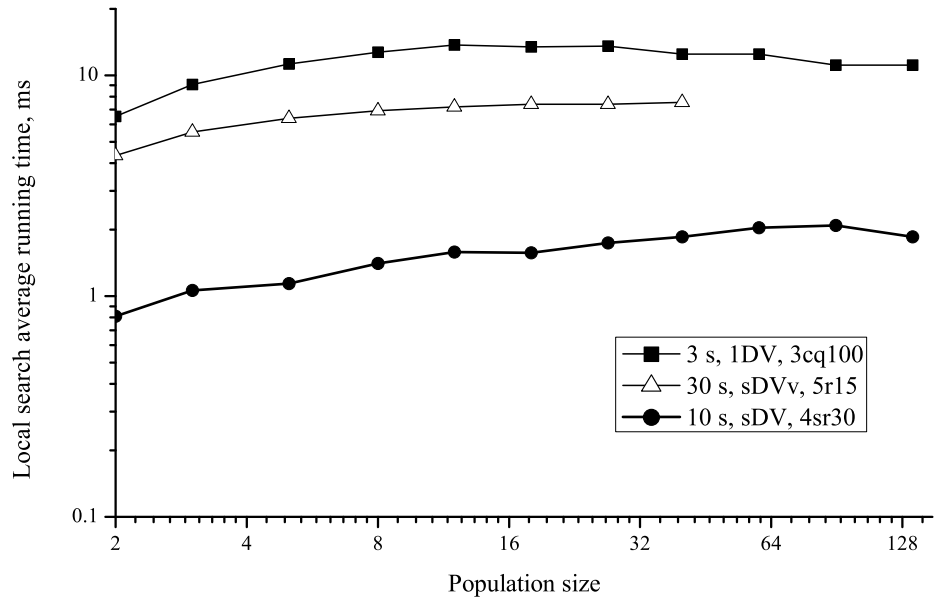

Figure 3: The average time required for one local search run depends only marginally on the proportion between the population size and the number of generations. These three lines correspond to three runs of our memetic algorithm. In every run we used different local search procedures (1DV, $s$DV and $s$DV$_\text{v}$, for details see Section 3.6) and different given times $\tau$ (3 s, 10 s and 30 s).

Note that all factors but the first one are hard to formalize. Next we will discuss relations between these factors.

Since we assume that almost only the local search consumes the processor time (see above), the computational platform affects only the local search procedure. Another parameter which greatly influences the local search performance is the problem instance; it is incorrect to discuss a local search performance without considering a particular instance.

Let $t$ be the average running time of the local search procedure applied to some solution of the given instance being run on the given computational platform. (Recall that this time stays almost constant during the algorithm run, see Figure 3.) Our idea is to use $t$ as the value which encapsulates the specifics of the instance, of the computational platform and of the local search procedure.

Definitely the local search and the instance are also related to the genetic operators and selection strategies, but we assume that this relation is not that important; our computational experience confirms this.

Hence, we can calculate the near-optimal population size $m_\text{opt} = f(t, \tau)$, and the rest of the factors are indirectly included into the function $f$ definition. Obviously $m_\text{opt}$ grows with the growth of $\tau$ and reduces with the growth of $t$. Let us use the following

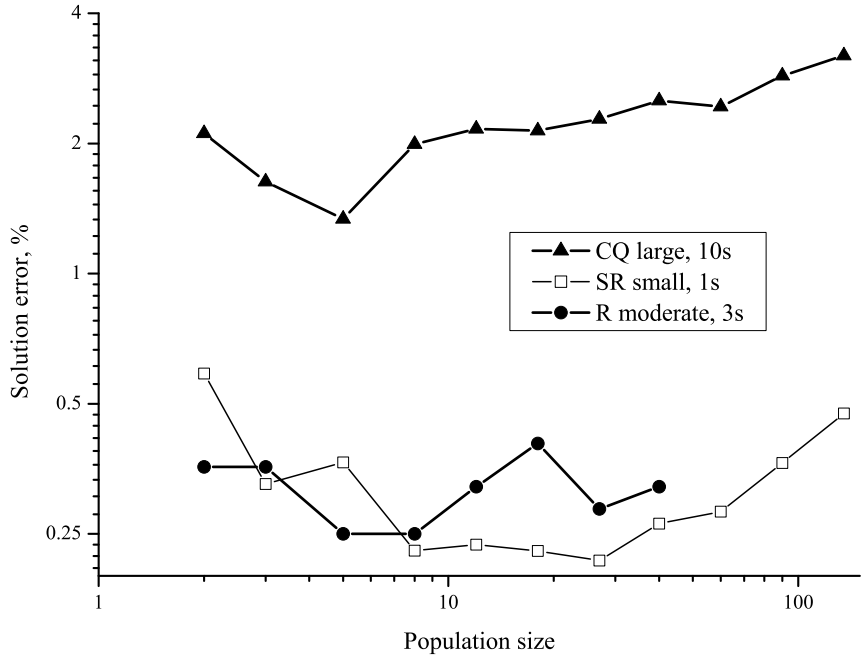

Figure 4: The solution quality significantly depends on the population size. For every instance, local search and given time, there exists some optimal population size. On this plot we show how the relative solution error depends on the population size for different types and sizes of instances (for detailed descriptions of the particular instance types, see Section 4).

flexible function for $m_{\text{opt}}$:

$$m_{\text{opt}}(\tau, t) = a \cdot \frac{\tau^b}{t^c}. \qquad (1)$$

The constants $a$, $b$ and $c$ are intended to reflect on the specifics of genetic operators and selection strategies. Observe that variation of $a$, $b$ and $c$ may significantly change the behavior of $m_{\text{opt}}$.

Since $a$, $b$ and $c$ are only related to the fixed parts of the algorithm, they should be adjusted before the algorithm's run, i.e., these parameters should be tuned (Eiben et al., 1999). However, the whole approach should be considered as a combination of parameter tuning and control since the time $t$ is obtained during the algorithm's run.

### 2.2  Choosing Constants $a$, $b$ and $c$

Our approach has two stages: tuning the constants $a$, $b$ and $c$ according to the algorithm structure, and finding the average running time $t$ of the local search procedure. Having all these values, we can calculate the near-optimal population size $m_{\text{opt}}$ according to (1) and run the algorithm.

This section discusses the first stage of our approach, i.e., tuning the constants $a$, $b$

and $c$. The next section discusses finding the value $t$.

The constants $a$, $b$ and $c$ in (1) should be selected to minimize the solution error for all combinations of local searches $\lambda$, instances $\phi$ and given times $\tau$ which are of interest. In practice this means that one should select a representative instance set $\Phi$, assign the most appropriate local search $\lambda = \lambda(\phi)$ for every instance $\phi \in \Phi$ and define several given times $\tau \in T$ which will be used in practice. Note that if $|T| = 1$, i.e., only one given time is required, then the number of constants in (1) can be reduced: $m_{\text{opt}}(t) = a/t^c$.

Let $A_{\text{MA}}(m, \lambda, \phi, \tau)$ be a solution obtained by the memetic algorithm for the population size $m$, local search $\lambda$, instance $\phi$ and given time $\tau$. Let $w(A)$ be the objective value of a solution $A$.

We need some measure of the memetic algorithm quality which reflects on the success of choosing a particular population size. This measure should not depend on the rest of the algorithm parameters, i.e., it should have similar values for all the solutions obtained for the best chosen population sizes whatever is the instance, the local search or the given time. Clearly one cannot use the relative solution error since its value hugely depends on the given time and other factors.

We propose using *scaled*[2] *solution errors* as follows. Let $w_{\min}(\lambda, \phi, \tau)$ and $w_{\max}(\lambda, \phi, \tau)$ be the minimum and the maximum objective values obtained for the given $\lambda$, $\phi$ and $\tau$:

$$w_{\min}(\lambda, \phi, \tau) = \min_m w(A_{\text{MA}}(m, \lambda, \phi, \tau)) \quad \text{and}$$

$$w_{\max}(\lambda, \phi, \tau) = \max_m w(A_{\text{MA}}(m, \lambda, \phi, \tau)) \ .$$

Then the scaled error $\epsilon(m, \lambda, \phi, \tau)$ of the solution $A_{\text{MA}}(m, \lambda, \phi, \tau)$ is calculated as follows:

$$\epsilon(m, \lambda, \phi, \tau) = \frac{w(A_{\text{MA}}(m, \lambda, \phi, \tau)) - w_{\min}(\lambda, \phi, \tau)}{w_{\max}(\lambda, \phi, \tau) - w_{\min}(\lambda, \phi, \tau)} \cdot 100\% \ .$$

In other words, the scaled solution error shows the position of the solution obtained for the given population size between the solutions obtained for the best and for the worst values of $m$. The scaled solution error is varied in $[0\%, 100\%]$; the smaller $\epsilon$, the better the solution. Note that this scaled error has some useful theoretical properties (Zemel, 1981).

Since all the scaled solution errors have comparable values, we can use the average for every combination of $\tau \in T$ and $\phi \in \Phi$ as an indicator of $m_{\text{opt}}$ function success:

$$\gamma = \overline{\epsilon\big(m_{\text{opt}}(\tau, t(\lambda, \phi)), \lambda, \phi, \tau\big)} \ . \tag{2}$$

(Note that we use $t(\lambda, \phi)$ because the average local search running time $t$ depends on the local search procedure $\lambda$ and the instance $\phi$; recall $\lambda = \lambda(\phi)$.) Obviously, $0\% \le \gamma \le 100\%$, and the smaller $\gamma$, the better $m_{\text{opt}}$.

The number of runs of the memetic algorithm required to find the best values of $a$, $b$ and $c$ can be huge[3] which makes the approach proposed in this paper unaffordable. For the purpose of decreasing the computation time we suggest the following dynamic programming technique.

1. Let $\Phi$ be the test bed and $T$ be the set of the given times we are going to use for our algorithm.

---

[2]Sometimes in the literature it is also called *differential*.
[3]Note that since memetic algorithms are stochastic, one should run every experiment several times in order to get a better precision.

2. For every instance $\phi \in \Phi$ set the most appropriate local search $\lambda = \lambda(\phi)$.

3. Let $M$ be the set of reasonable population sizes. One can even reduce it by removing, e.g., all odd values from $M$, or leaving only certain values, e.g., $M = \{2, 4, 8, 16, \ldots\}$.

4. Calculate and save $e(m, \lambda(\phi), \phi, \tau)$ for every $m \in M$, $\phi \in \Phi$ and $\tau \in T$.

5. Measure and save $t(\lambda(\phi), \phi)$ for every $\phi$. For this purpose run the local search $\lambda(\phi)$ after a construction heuristic.

6. Now for every combination of $a$, $b$ and $c$ compute $\gamma$ according to (2); every time the relative solution error $e(m, \lambda(\phi), \phi, \tau)$ is required, find $m' \in M$ which is the closest one to $m$ and use the corresponding precalculated value. The discretization of $a$, $b$ and $c$ should be chosen according to available resources.

7. Fix the combination of $a$, $b$ and $c$ which minimizes $\gamma$. This finishes the tuning process.

### 2.3 Finding Local Search Average Running Time $t$

In order to calculate the near-optimal population size $m_{\text{opt}}$ according to (1), we need to find $t$ at the beginning of the memetic algorithm run. Recall that the value $t$ is the average running time of the local search procedure applied to some solutions of the given instance. Definitely this value significantly depends on the particular solutions. However, the solutions in a memetic algorithm are permanently perturbed and, thus, they are always moved out from the local minima before the local search is applied. This guaranties some uniformity in the improvement process during the whole algorithm. Hence, we are able to measure the time $t$ at any point.

Our algorithm produces and immediately improves the solutions for the first generation until $m_1 \leq m_{\text{opt}}(\tau, t_{\text{cur}}/m_1)$, where $m_1$ is the number of already produced solutions, $\tau$ is the time given to the whole memetic algorithm, $t_{\text{cur}}$ is the time already spent to generate solutions for the first generation and $m_{\text{opt}}(\tau, t)$ is the population size calculated according to (1). When the first generation is produced, the size of the population for all further generations is set to $m = m_{\text{opt}}(\tau, t_{\text{cur}}/m_1)$.

## 3 Case Study: Algorithm for Multidimensional Assignment Problem

As a case study for the population sizing proposed in Section 2 we decided to use the Multidimensional Assignment Problem (MAP); for the problem review see Section 1.

### 3.1 Main Algorithm Scheme

While the general scheme of a typical memetic algorithm (see Figure 1) is quite common for all memetic algorithms, the set of genetic operators and the way they are applied can vary significantly. In this paper we use quite a typical (see, e.g., Krasnogor and Smith (2008)) procedure to obtain the next generation:

$$g^{i+1} = selection\left(\{g_1^i\} \cup mutation\left(g^i \setminus \{g_1^i\}\right) \cup crossover\left(g^i\right)\right), \qquad (3)$$

where $g^k$ is the $k$th generation and $g_1^k$ is the best assignment in the $k$th generation. For a set of assignments $G$ the function $selection(G)$ simply returns $m_{i+1}$ best distinct assignments among them, where $m_k$ is the size of the $k$th generation (if the number of distinct assignments in $G$ is less than $m_{i+1}$, $selection$ returns all the distinct assignments

New Approach to Population Sizing: Case Study for MAP

and updates the value of $m_{i+1}$ accordingly). Note that the assignment $g_1^i$ avoids the *mutation* thus preserving the currently best result. The function $mutation(G)$ is defined as follows:

$$mutation(G) = \bigcup_{g \in G} \begin{cases} LocalSearch(perturb(g, \mu_m)) & \text{if } r < p_m \\ g & \text{otherwise} \end{cases} \quad (4)$$

where $r \in [0, 1]$ is chosen randomly every time and the constants $p_m = 0.5$ and $\mu_m = 0.1$ define the probability and the strength of mutation operator respectively. The function $crossover(G)$ is calculated as follows:

$$crossover(G) = \bigcup_{j=1}^{(l \cdot m_{i+1} - m_i)/2} LocalSearch(crossover(u_j, v_j)) \quad (5)$$

where $u_j$ and $v_j$ are assignments from $G$ randomly selected for every $j = 1, 2, \ldots, (l \cdot m_{i+1} - m_i)/2$ and $l = 3$ defines ratio between the produced and selected for the next generation solutions. The functions $crossover(x, y)$, $perturb(x, \mu)$ and $LocalSearch(x)$ are discussed below.

### 3.2 Coding

Coding is a way of representing a solution as a sequence of atom values such as boolean values or numbers; genetic operators are applied to such sequences. Good coding should meet the following requirements:

- Coding $code(x)$ should be invertible, i.e., there should exist a decoding procedure $decode$ such that $decode(code(x)) = x$ for any feasible solution $x$.

- Evaluation of the quality (fitness function) of a coded solution should be fast.

- Every fragment of the coded solution should refer to just a part of the whole solution, so that a small change in the coded sequence should not change the whole solution.

- It should be relatively easy to design algorithms for random modification of a solution (mutation) and for combination of two solutions (crossover) which produce feasible solutions.

Huang and Lim (2006) use a local search procedure which, given first two dimensions of an assignment, determines the third dimension (recall that the algorithm by Huang and Lim (2006) is designed only for 3-AP). Since the first dimension can always be fixed without any loss of generality (see Section 1), one needs to store only the second dimension of an assignment. Unfortunately, this coding requires a specific local search and is robust for 3-AP only. We use a different coding; a vector of an assignment is considered as an atom in our algorithm and, thus, a coded assignment is just a list of its vectors. The vectors are always stored in the first coordinate ascending order, e.g., an assignment consisting of vectors $(2, 1, 1)$, $(4, 4, 2)$, $(3, 2, 3)$ and $(1, 3, 4)$ (see Fig. 2) would be represented as

$$(1, 3, 4), (2, 1, 1), (3, 2, 3), (4, 4, 2).$$

Two assignments are considered equal if they have equal codes.

G. Gutin, D. Karapetyan

### 3.3 First Generation

As it was shown by Gutin and Karapetyan (2009a) (and we also confirmed it empirically by testing our memetic algorithm with construction heuristics described in (Karapetyan et al., 2009)), it is beneficial to start any MAP local search or metaheuristic from a Greedy construction heuristic. Thus, we start from running Greedy (we use the same implementation as in (Gutin and Karapetyan, 2009a)) and then perturb it using our $perturb$ procedure (see Section 3.5) to obtain every item of the first generation:

$$g_j^1 = LocalSearch(perturb(greedy, \mu_f)),$$

where $greedy$ is an assignment constructed by Greedy and $\mu_f = 0.2$ is the perturbation strength coefficient. Since $perturb$ performs a random modification, it guarantees some diversity in the first generation.

The number of assignments to be produced for the first generation is discussed in Section 2.3.

### 3.4 Crossover

A typical crossover operator combines two solutions, parents, to produce two new solutions, children. Crossover is the main genetic operator, i.e., it is the source of a genetic algorithm strength. Due to the selection operator, solutions consisting of 'successful' fragments are spread wider than others and that is why, if both parents have some similar fragments, these fragments are assumed to be 'successful' and should be copied without any change to the children solutions. Other parts of the solution can be randomly mixed and modified though they should not be totally destroyed.

The one-point crossover is the simplest example of a crossover; it produces two children $x'$ and $y'$ from two parents $x$ and $y$ as follows: $x_i' = x_i$ and $y_i' = y_i$ for every $i = 1, 2, \ldots, k$, and $x_i' = y_i$ and $y_i' = x_i$ for every $i = k+1, k+2, \ldots, n$, where $k \in \{1, 2, \ldots, n-1\}$ is chosen randomly. One can see that if $x_i = y_i$ for some $i$, then the corresponding values in the children sequences will be preserved: $x_i' = y_i' = x_i = y_i$.

However, the one-point and some other standard crossovers do not preserve feasibility of MAP assignments since not every sequence of vectors can be decoded into a feasible assignment. We propose a special crossover operator. Let $x$ and $y$ be the parent assignments and $x'$ and $y'$ be the child assignments. First, we retrieve equal vectors in the parent assignments and initialize both children with this set of vectors:

$$x' = y' = x \cap y\,.$$

Let $k = |x \cap y|$, i.e., the number of equal vectors in the parent assignments, $p = x \setminus x'$ and $q = y \setminus y'$, where $p$ and $q$ are ordered sets. Let $\pi$ and $\omega$ be random permutations of size $n - k$. Let $r$ be an ordered set of random values uniformly distributed in $[0, 1]$. For every $j = 1, 2, \ldots, n-k$ the crossover sets

$$x' = x' \cup \begin{cases} p_{\pi(j)} & \text{if } r_j < 0.8 \\ q_{\omega(j)} & \text{otherwise} \end{cases} \quad \text{and} \quad y' = y' \cup \begin{cases} q_{\omega(j)} & \text{if } r_j < 0.8 \\ p_{\pi(j)} & \text{otherwise} \end{cases}.$$

Since this procedure can yield infeasible assignments, it requires additional correction of the child solutions. For this purpose, the following is performed for every dimension $d = 1, 2, \ldots, s$ and for every child assignment $c$. For every $i$ such that $\exists j < i : c_d^j = c_d^i$ set $c_d^i = r$ where $r \in \{1, 2, \ldots, n\} \setminus \{c_d^1, c_d^2, \ldots, c_d^n\}$ is chosen randomly. In the end of the correction procedure, sort the assignment vectors in the ascending order of the first coordinates (see Section 3.2).

In other words, our crossover copies all equal vectors from the parent assignments to the child ones. Then it copies the rest of the vectors; every time it chooses randomly a pair of vectors, one from the first parent and one from the second one. Then it adds this pair of vectors either to the first and to the second child respectively (probability 80%) or to the second and to the first child respectively (probability 20%). Since the obtained child assignments can be infeasible, the crossover corrects each one; for every dimension of every child it replaces all duplicate coordinates with randomly chosen correct ones, i.e., with the coordinates which are not currently used for that dimension.

Note that (5) requires $l \cdot m_{i+1} - m_i$ to be even. If $m_{i+1} = m_i = m_o(\tau, t)$ then $l \cdot m_{i+1} - m_i$ is always even (recall that $l = 3$). However, the size of the population is not guaranteed and, hence, $l \cdot m_{i+1} - m_i = (l - 1) \cdot m$ may take odd values. To resolve this issue, we remove the worst solution from the $i$th generation if $l \cdot m_{i+1} - m_i$ appears to be odd.

We also tried the crossover operator used in (Huang and Lim, 2006) but it appeared to be less efficient than the one proposed here.

### 3.5 Perturbation Algorithm

The perturbation procedure $perturb(x, \mu)$ is intended to modify randomly an assignment $x$, where the parameter $\mu$ defines how strong is the perturbation. In our memetic algorithm, perturbation is used to produce the first generation and to mutate assignments from the previous generation when producing the next generation.

Our perturbation procedure $perturb(x, \mu)$ performs $\lceil n\mu/2 \rceil$ random swaps. In particular, each swap randomly selects two vectors and some dimension and then swaps the corresponding coordinates: swap $x_u^d$ and $x_v^d$, where $u, v \in \{1, 2, \ldots, n\}$ and $d \in \{1, 2, \ldots, s\}$ are chosen randomly; repeat the procedure $\lceil n\mu/2 \rceil$ times. For example, if $\mu = 1$, our perturbation procedure modifies up to $n$ vectors in the given assignment.

### 3.6 Local Search Procedure

An extensive study of a number of local search heuristics for MAP can be found in (Gutin and Karapetyan, 2009a); the paper includes both fast and slow algorithms. It also shows that a combination of two heuristics can yield a heuristic superior to the original ones.

The following heuristics were considered as candidates for the local search procedure for our memetic algorithm (we provide only a brief description of every heuristic here; full descriptions can be found in (Gutin and Karapetyan, 2009a)):

- 1DV, 2DV and $s$DV are dimensionwise (Gutin and Karapetyan, 2009a) local searches. On every iteration, they fix some dimensions while the other dimensions are grouped together. The problem of optimal matching the fixed and unfixed parts of the assignment vectors can be represented as 2-AP which is solvable in the polynomial time. 1DV, 2DV and $s$DV fix up to one, two and $s$ dimensions on every iteration, respectively.

- 2-opt (3-opt) is a simple heuristic that selects the best of all possible recombinations for every pair (triple) of vectors in the assignment. 2-opt is known as a very fast but poor quality heuristic. 3-opt is a high quality but slow local search which has no application as a stand-alone heuristic but is useful in a combination with dimensionwise heuristics (Gutin and Karapetyan, 2009a).

- v-opt is an extension of the Variable Depth Interchange heuristic which was initially proposed in (Balas and Saltzman, 1991) for 3-AP. Like 2-opt, v-opt considers

  recombinations of vector pairs, however the objective and the enumeration order in v-opt are totally different.

- $1DV_2$, $2DV_2$, $sDV_v$ and $sDV_3$ are combinations of 1DV, 2DV or $s$DV with 2-opt, 3-opt or v-opt. A variable local search is exploited here; the first and the second heuristics are applied sequentially to the given assignment until no further improvement can be obtained.

Results for 3-opt and v-opt as a local search for our memetic algorithm are not provided in this paper since they did not show any promising results in our experiments; Gutin and Karapetyan (2009a) also indicate them to be inefficient heuristics.

Gutin and Karapetyan (2009a) propose a division of instances into two groups: instances with independent weights and instances with decomposable weights. The weight matrices of the instances with independent weights have no structure, i.e., there is no correlation between weights $w(u)$ and $w(v)$ even if the vectors $u$ and $v$ are different in only one coordinate. In contrast, the weights of the instances with decomposable weights are defined using the graph formulation of MAP (see Section 1) and have the following structure:

$$w(e) = f\left(d^{1,2}_{e_1,e_2}, d^{1,3}_{e_1,e_3}, \ldots, d^{s-1,s}_{e_{s-1},e_s}\right) , \qquad (6)$$

where matrix $d^{i,j}$ define weights of the edges between sets $X_i$ and $X_j$, and $f$ is some function. Most of the instances which have some practical interest and which do not belong to the group of independent weight instances can be represented as instances with decomposable weights, see, e.g., Clique and SquareRoot instance families in Section 4.

It is known that even for a fixed optimization problem there is no local search procedure which would be the best choice for all types of instances (Krasnogor and Smith, 2001, 2005). Splitting all the MAP instances into two groups, namely instances with independent and decomposable weights, gives us a formal way to use appropriate local searches for every instance. In particular, it was shown by Gutin and Karapetyan (2009a) that the instances with independent weights are better solvable by $sDV_v$ while the dimensionwise heuristics are the best choice for the instances with decomposable weights.

Table 1 presents a comparison of the results of our memetic algorithm based on the local search procedures discussed above. The time given for every run of the algorithm is 3 seconds. The table reports the relative solution error for every instance and every considered algorithm. The column 'best' shows the best known solution for each instance.

One can see that the outcomes of (Gutin and Karapetyan, 2009a) are repeated here, i.e., for the Random instances (see Section 4) $sDV_v$ provides clearly the best performance; for the instances with decomposable weights, i.e., for the Clique and SquareRoot instances, the fast heuristics 1DV, 2DV, $s$DV, $1DV_2$ and $2DV_2$ perform better than others in almost every experiment, and $s$DV shows the best average result among them (though in Table 1 2DV slightly outperforms it, for other given times $s$DV shows the best results).

Thereby, in what follows we use $sDV_v$ as a local search for the instances with independent weights and $s$DV for the instances with decomposable weights.

### 3.7 Population Size Adjustment

The constants $a$, $b$ and $c$ were selected to minimize $\gamma$ (see Section 2.1); as an instance set $\Phi$ we used the full test bed (see Section 4), the given times were $T = \{1\text{s}, 3\text{s}, 10\text{s}, 30\text{s}, 100\text{s}\}$, the generation sizes were $M = \{2, 3, 5, 8, 12, 18, 27, 40, 60, 90, 135\}$ and local search $\lambda(\phi)$ was selected according to Section 3.6. The best value of $\gamma = 13\%$ was obtained for $a = 0.08$, $b = 0.35$ and $c = 0.85$ (see (1)). Note that these values are not a compromize and present minima for every separate instance set and given time. Observe also that fixing $m$ to some value leads to $\gamma > 19\%$ for the same set of instances, local searches and given times.

Slight variations of the constants $a$, $b$ and $c$ do not influence the performance of the algorithm significantly. Moreover, there exist some other values for these parameters which also yield good results. The values of the constants should not be adjusted for every computational platform.

## 4 Test Bed

In this section we discuss instance families used for experimental evaluation of our memetic algorithm. As it was mentioned above, we use two types of instances: instances with independent weights (Random) and instances with decomposable weights (Clique, SquareRoot, Geometric and Product).

The Random instances simply assign a uniformly distributed random weight to every vector $e \in X$. The weight was chosen from $\{1, 2, \ldots, 100\}$ in our experiments. We believe that Random instances are of a small practical interest and we included them in the test bed because they are widely used in the literature and also because of their theoretical properties (Grundel et al., 2004; Gutin and Karapetyan, 2009c).

Initially we have also considered pseudo-random instances with predefined optimal solutions (Grundel and Pardalos, 2005). However, the generator of these instances has the exponential time complexity and the time required to generate the instances of this type of appropriate size for our test bed is beyond any reasonable value.

The Clique and SquareRoot instance families have decomposable weights (see (6)) and, thus, they are defined for weighted $s$-partite graphs $G = (X_1 \cup X_2 \cup \ldots \cup X_s, E)$. Weight $w(e)$ of every edge $e \in E$ was initialized independently and randomly in our experiments; $w(e)$ was chosen uniformly from $\{1, 2, \ldots, 100\}$. Let $C$ be a clique in $G$ and let $E_C$ be the set of edges induced by this clique. Then the weight of a vector corresponding to the clique $C$ is calculated as follows for the Clique and SquareRoot instances respectively:

$$w_{\text{CQ}}(E_C) = \sum_{e \in E_C} w(e) \quad \text{and}$$

$$w_{\text{SR}}(E_C) = \sqrt{\sum_{e \in E_C} w(e)^2} \,,$$

i.e., in the case of SquareRoot, the objective is not only to minimize the considered weights, like it is for Clique, but also to keep all the weights not too large.

A special case of Clique is Geometric instance family. In Geometric, the sets $X_1$, $X_2$, $\ldots$, $X_s$ (see Section 1) correspond to $s$ sets of points in a Euclidean space, and the distance between two points $u \in X_i$ and $v \in X_j$ is defined as the Euclidean distance; we consider two dimensional Euclidean space:

$$d_{\text{g}}(u, v) = \sqrt{(u_x - v_x)^2 + (u_y - v_y)^2} \,.$$

It is proven (Spieksma and Woeginger, 1996) that the Geometric instances are NP-hard to solve for $s = 3$ and, thus, Geometric is NP-hard for every $s \geq 3$.

Product is another NP-hard (Burkard et al., 1996b) instance family with decomposable weights. A weight of a vector $e$ in Product is defined as follows:

$$w_{\text{P}}(e) = \prod_{j=1}^{s} a_{e_j}^{j} ,$$

where $a^j$ is an array of $n$ values, each randomly selected from $\{1, 2, \ldots, 100\}$.

Our test bed includes instances of 3-AP, 4-AP, 5-AP and 6-AP; for every number of dimensions three sizes $n$ are used which correspond to small, moderate and large instances. For every combination of $s$ and $n$ 50 instances (10 Random, 10 Clique, 10 Square-Root, 10 Geometric and 10 Product instances) are included into the test bed. Thereby, we produced 10 different instances for every combination of $s$, $n$ and instance family and, thus, every number reported in the tables in Section 5 is average among 10 runs. We use standard Miscrosoft .NET random generator (Microsoft, 2008) which is based on the Donald E. Knuth's subtractive random number generator algorithm (Knuth, 1981). As a seed of the random number sequences for all the instance types we use the following number: $seed = s + n + i$, where $i \in \{1, 2, \ldots, 10\}$ is the index of the instance.

## 5 Experimental Evaluation

Three metaheuristics were compared in our experiments:

- An extended version of the memetic algorithm by Huang and Lim (2006) (HL).

- An extended version of the simulated annealing algorithm by Clemons et al. (2004) (SA).

- Our memetic algorithm (GK).

All the heuristics are implemented in Visual C++ and evaluated on a platform based on AMD Athlon 64 X2 3.0 GHz processor. The implementations as well as the test bed generator and the best known assignments are available on the web (Karapetyan, 2009).

### 5.1 HL Heuristic

For the purpose of comparison, the Huang and Lim's memetic algorithm was extended as follows:

- The coded assignment contains not only the second dimension but it stores sequentially all the dimensions except the first and the last ones, i.e., an assignment $\{e^1, e^2, \ldots, e^s\}$ is represented as $e_2^1, e_2^2, \ldots, e_2^n, e_3^1, e_3^2, \ldots, e_3^n, \ldots, e_{s-1}^1, e_{s-1}^2, \ldots, e_{s-1}^n$ ($e_1^i = i$ for each $i$ and $e_s^i$ can be chosen in an optimal way by solving an AP, see Section 3.2).

- The local search heuristic, that was initially designed for 3-AP, is extended to 1DV as described in (Gutin and Karapetyan, 2009a).

- The crossover, proposed in (Huang and Lim, 2006), is applied separately to every dimension (except the first and the last ones) since it was designed for one dimension only (recall that the memetic algorithm from (Huang and Lim, 2006) stores only the second dimension of an assignment, see Section 3.2).

- The termination criterion is replaced with a time check; the algorithm terminates when the given time is elapsed.

Our computational experience show that the solution quality of our implementation of the Huang and Lim's heuristic is similar to the results reported in (Huang and Lim, 2006) and the running time is reasonably larger because of the extension for $s > 3$.

### 5.2 SA Heuristic

The Simulated Annealing heuristic by Clemons et al. (2004) was initially proposed for arbitrary number of dimensions. We reimplemented it and our computational experience show that both the solution quality and the running times[4] of our implementation of the Simulated Annealing heuristic are similar to the results reported in (Clemons et al., 2004).

For the purpose of comparison to other heuristics we needed to fit SA for using a predefined running time. We tried two strategies:

- An adaptive cooling ratio $R$ (see Clemons et al. (2004)). The value $R$ is updated before each change of the temperature as follows:

$$R = \sqrt[m]{\frac{0.1}{T}} \quad \text{and} \quad f = (\tau - t_e) \cdot \frac{i}{t_e} \,,$$

where $T$ is the current temperature (see Clemons et al. (2004)), $t_e$ is the elapsed time, $\tau$ is the given time and $f$ is the expected number of further iterations which is calculated according to the number $i$ of already finished iterations.

- An adaptive number of local search iterations $NUM_{\max}$ (see Clemons et al. (2004)). The value $NUM_{\max}$ is updated before each change of the temperature as follows:

$$NUM_{\max} = (\tau - t_e) \cdot \frac{c}{t_e} \cdot \frac{1}{I - i} \,,$$

where $t_e$ is the elapsed time, $\tau$ is the given time, $c$ is the total number of local search iterations already performed, $i$ is the number of the algorithm iterations already performed and $I$ is the number of algorithm iterations to be performed. Since the cooling ratio $R$ as well as the initial and the final temperatures $T_{\text{start}}$ and $T_{\text{final}}$ are fixed, the number $I$ of iterations of the algorithm is also fixed:

$$I = \log_R \frac{T_{\text{final}}}{T_{\text{initial}}} \,.$$

For both adaptations the algorithm terminates if the given time is elapsed: $t \geq \tau$.

Both adaptations yielded competitive algorithms though according to our experimental evaluation the second adaption which varies the number of local search iterations appears to be more efficient. One can assume that the best adaptation should vary both the cooling ratio and the number of local search iterations but this is a subject for another research. Hence, in what follows the SA algorithm refers to the extension with the adaptive number of local search iterations.

---

[4]In our experiments, the running times of the heuristic were always approximately 20 times smaller than the results reported in (Clemons et al., 2004) which can be explained by a difference in the computational platforms.

### 5.3 Experiment Results

The main results are reported in Tables 2 and 3; in these tables, we compare our algorithm (GK) to the Simulated Annealing heuristic (SA) and the memetic algorithm by Huang and Lim (HL). The comparison is performed for the following given times $\tau$: 0.3 s, 1 s, 3 s, 10 s, 30 s, 100 s and 300 s. Every entry of these tables contains the relative solution error averaged for 10 instances of some fixed type and size but of different seed values (see Section 4 for details); we did not repeat every experiment several times which is typical for stochastic algorithms. The value of the relative solution error $e(A)$ is calculated as follows

$$e(A) = \left(\frac{w(A)}{w(A_{\text{best}})} - 1\right) \cdot 100\% . \tag{7}$$

where $A$ is the obtained solution and $A_{\text{best}}$ is the best known solution[5].

The name of an instance consists of three parts: the number of dimensions $s$, the type of the instance ('r' for Random, 'cq' for Clique and 'sr' for SquareRoot) and the size $n$ of the instance. The results for Product and Geometric instances were excluded from Tables 1, 2 and 3 because even stand alone local searches used in our memetic algorithm are able to solve Geometric instances to optimality and Product instances to less than 0.04% over optimality[6]. Similar result were reported in (Gutin and Karapetyan, 2009a).

The average values for different instance families, numbers of dimensions and instance sizes are provided at the bottom of each table. The best among HL, SA and GK results are underlined in every row for every particular given time.

One can see that GK clearly outperforms both SA and HL for all the given times. Moreover, GK is not worse than the other heuristics in every experiment which proves its flexibility and robustness. A two-sided paired $t$-test confirms statistical difference even between GK with $\tau = 1$ s and HL with $\tau = 100$ s because the $p$-value in this case was less than $0.0001$ for both instances with independent and decomposable weights. This shows that HL is not able to use large time efficiently.

The solution quality of GK significantly depends on the given time: for the instances with both independent and decomposable weights a three times increase of the running time improves the solution quality approximately 1.2 to 2 times for the large and small $\tau$, respectively. Recall that the approach proposed in this paper to select the most appropriate population size reduces $\gamma$ more than 1.5 times (see Section 2.1) and, hence, it would take roughly 1.5 to 10 times more time to get the same solution quality for a memetic algorithm with a fixed population size[7].

It is worth noting that we experimented with different values of the GK algorithm parameters such as $\mu_f$, $\mu_m$, $p_m$, $l$, etc. and concluded that small variations of these values do not significantly influence the algorithm performance.

For the instances with independent weights all the algorithms perform better for the large instances rather than for the small ones. One can explain it by showing that the number of vectors of the minimal weight in Random is proportional to $n^s$ while the number of vectors in an assignment is $n$ and, thus, the number of global minima increases with the increase of $n$ (Grundel et al., 2004; Gutin and Karapetyan, 2009a). In contrast, the instances with decomposable weights become harder with the growth of $n$.

---

[5] The best known solutions were obtained during our experiments with different heuristics and the corresponding weights can be found in Table 1. For the Random instances we actually know the optimal objective values; it is proven for large values of $n$ that a Random instance has a solution of the minimal possible weight (Gutin and Karapetyan, 2009c); since we obtained the minimal possible solutions for every Random instance in our experiments, we can extend the results of Gutin and Karapetyan (2009c) to all the Random instances in our test bed.

[6] We believe that the best known solutions for both Geometric and Product instances are optimal but we are not able to verify it.

[7] Note that $\gamma$ is not just the average for the solution errors and, thus, these calculations are very approximate.

Since the HL heuristic uses 1DV local search, it performs quite well for the instances with decomposable weights and yields solutions of poor quality for the instances with independent weights. Due to the fixed population size, it does not manage to solve some large instances in short times which results in huge solution errors reported in Table 2 for the instances 3cq70, 3sr70, 3cq100 and 3sr100. HL was initially designed for 3-AP and tested on small instances (Huang and Lim, 2006) and, hence, it performs better for the instances with small $s$ and $n$.

The SA heuristic is less successful than the others; for both instances with independent and decomposable weights it is worse than both HL and GK in almost every experiment. The solution quality of SA improves quite slowly with the increase of the running time; it seems that SA would not be able to significantly improve the solution quality even if it is given much larger time.

## 6 Conclusion

In this paper, we propose a new approach to population sizing in memetic algorithms. As a case study, we designed and evaluated a memetic algorithm for the Multidimensional Assignment Problem. Our experiments have confirmed that the proposed population sizing leads to an outstanding flexibility of the algorithm. Indeed, it was able to perform efficiently for a wide range of instances, being given from 0.3 to 300 seconds of the running time and with totally different local search procedures. As an evidence of its efficiency, we compared it with two other metaheuristics proposed in the literature and concluded that our algorithm clearly outperforms the other heuristics with no exception. Moreover, the difference in the solution quality of our memetic algorithm (GK) and the previous state-of-the-art memetic algorithm (HL) continuously grows with the increase of the given time which confirms that GK is much more flexible than HL.

The main factors influencing the performance of a memetic algorithm are running time, computational platform, problem instance, local search procedure, population size and genetic operators. We did not focus on the genetic operators investigation in this research; however we believe that the operators used in our algorithm are well fitted since our attempts to improve the algorithm results by changing the operators have failed. The local search procedure and the population size are varied according to the problem instance; after an extensive study of the local searches, we show that there are two totally different cases of MAP, and for these cases one should use different local search procedures. Since these local searches have very different running times, the memetic algorithm should adapt for them. This is done by using the adjustable population size which is a function of the average running time of the local search. Thereby, the average running time of the local search encapsulates not only the local search specifics but also the specifics of the instance and the computational platform performance. Since the algorithm is self-adjustable, the running time can be used as a parameter responsible for the 'solution quality'/'running time' balance and, thus, the population size should also depend on the given time.

The adjustable population size requires several constants to be tuned prior to using the algorithm; we proposed a procedure to find the optimal values of these constants.

In conclusion we note that choosing the most appropriate population size is crucial

G. Gutin, D. Karapetyan

for the performance of a memetic algorithm. Our approach to calculate the population size according to the average running time of the local search and the time given to the whole algorithm, used to perform well for a large variation of the instances and given times and for two totally different local searches. Observe, however, that the whole discussion of the population sizing does not involve any MAP specifics and, hence, we can conclude that the obtained results can be extended to any hard optimization problem.

Further research is required to evaluate the proposed approach in application to other hard combinatorial optimization problems. It is also an interesting question if changing the population size during the algorithm's run can further improve the results.

Table 1: Memetic algorithms based on different local search comparison. The given time is 3 s.

| | | Relative solution error, % | | | | | | | |
|---|---|---|---|---|---|---|---|---|---|
| Inst. | Best | 2-opt | 1DV | 2DV | $sDV$ | $1DV_2$ | $2DV_2$ | $sDV_3$ | $sDV_v$ |
| 3r40 | 40.0 | 122.00 | 26.75 | 30.00 | 27.25 | 32.25 | 32.50 | 32.00 | 6.25 |
| 3r70 | 70.0 | 102.71 | 11.43 | 11.14 | 11.57 | 11.71 | 11.57 | 15.00 | 0.71 |
| 3r100 | 100.0 | 83.90 | 3.00 | 3.20 | 3.10 | 3.30 | 3.10 | 5.80 | 0.00 |
| 4r20 | 20.0 | 68.00 | 46.00 | 28.00 | 29.50 | 39.50 | 32.00 | 17.50 | 0.00 |
| 4r30 | 30.0 | 73.00 | 31.00 | 23.67 | 23.67 | 27.00 | 21.67 | 14.33 | 0.00 |
| 4r40 | 40.0 | 73.50 | 24.00 | 15.25 | 15.00 | 23.00 | 15.75 | 11.25 | 0.00 |
| 5r15 | 15.0 | 36.67 | 39.33 | 19.33 | 16.67 | 22.00 | 21.33 | 8.00 | 0.00 |
| 5r18 | 18.0 | 40.56 | 37.78 | 20.56 | 19.44 | 26.11 | 18.89 | 2.78 | 0.00 |
| 5r25 | 25.0 | 40.40 | 34.00 | 16.80 | 16.80 | 25.60 | 18.40 | 3.60 | 0.00 |
| 6r12 | 12.0 | 10.00 | 39.17 | 15.83 | 10.00 | 14.17 | 13.33 | 0.83 | 0.00 |
| 6r15 | 15.0 | 22.00 | 45.33 | 16.67 | 11.33 | 18.67 | 13.33 | 0.00 | 0.00 |
| 6r18 | 18.0 | 23.89 | 37.22 | 18.33 | 10.00 | 17.22 | 12.78 | 0.00 | 0.00 |
| All avg. | | 58.05 | 31.25 | 18.23 | 16.19 | 21.71 | 17.89 | 9.26 | 0.58 |
| 3-AP avg. | | 102.87 | 13.73 | 14.78 | 13.97 | 15.75 | 15.72 | 17.60 | 2.32 |
| 4-AP avg. | | 71.50 | 33.67 | 22.31 | 22.72 | 29.83 | 23.14 | 14.36 | 0.00 |
| 5-AP avg. | | 39.21 | 37.04 | 18.90 | 17.64 | 24.57 | 19.54 | 4.79 | 0.00 |
| 6-AP avg. | | 18.63 | 40.57 | 16.94 | 10.44 | 16.69 | 13.15 | 0.28 | 0.00 |
| Small avg. | | 59.17 | 37.81 | 23.29 | 20.85 | 26.98 | 24.79 | 14.58 | 1.56 |
| Moderate avg. | | 59.57 | 31.38 | 18.01 | 16.50 | 20.87 | 16.37 | 8.03 | 0.18 |
| Large avg. | | 55.42 | 24.56 | 13.40 | 11.23 | 17.28 | 12.51 | 5.16 | 0.00 |
| 3cq40 | 939.9 | 12.45 | 0.05 | 0.01 | 0.10 | 0.04 | 0.11 | 2.60 | 0.31 |
| 3sr40 | 610.6 | 15.39 | 0.05 | 0.23 | 0.07 | 0.23 | 0.25 | 2.46 | 0.23 |
| 3cq70 | 1158.4 | 37.92 | 3.84 | 3.98 | 3.43 | 4.72 | 4.63 | 10.50 | 5.94 |
| 3sr70 | 737.1 | 44.15 | 4.79 | 5.28 | 5.70 | 4.94 | 5.06 | 14.30 | 6.46 |
| 3cq100 | 1368.1 | 47.09 | 8.19 | 7.92 | 8.29 | 8.61 | 8.82 | 15.04 | 10.55 |
| 3sr100 | 866.3 | 46.02 | 7.92 | 7.77 | 7.61 | 8.50 | 8.48 | 14.71 | 11.06 |
| 4cq20 | 1901.8 | 0.27 | 0.01 | 0.02 | 0.03 | 0.08 | 0.06 | 1.16 | 0.27 |
| 4sr20 | 929.3 | 0.40 | 0.01 | 0.12 | 0.03 | 0.14 | 0.03 | 0.85 | 0.36 |
| 4cq30 | 2281.9 | 5.53 | 0.41 | 0.69 | 0.69 | 0.67 | 0.73 | 5.26 | 1.77 |
| 4sr30 | 535.1 | 20.15 | 5.05 | 2.15 | 2.32 | 4.20 | 2.39 | 9.81 | 5.12 |
| 4cq40 | 2606.3 | 14.53 | 2.98 | 1.96 | 2.47 | 2.90 | 3.49 | 9.04 | 6.85 |
| 4sr40 | 1271.4 | 19.85 | 5.86 | 5.15 | 4.41 | 5.33 | 4.62 | 13.43 | 9.32 |
| 5cq15 | 3110.7 | 0.01 | 0.00 | 0.00 | 0.00 | 0.00 | 0.00 | 1.53 | 0.01 |
| 5sr15 | 1203.9 | 0.24 | 0.02 | 0.00 | 0.02 | 0.04 | 0.00 | 2.22 | 0.10 |
| 5cq18 | 3458.6 | 0.30 | 0.00 | 0.04 | 0.04 | 0.02 | 0.00 | 2.90 | 0.30 |
| 5sr18 | 504.9 | 3.72 | 1.47 | 0.04 | 0.00 | 0.28 | 0.24 | 4.12 | 0.61 |
| 5cq25 | 4192.7 | 4.03 | 0.25 | 0.54 | 0.54 | 0.86 | 0.87 | 6.82 | 2.71 |
| 5sr25 | 1627.5 | 4.68 | 0.44 | 1.04 | 1.14 | 0.58 | 1.27 | 8.31 | 3.90 |
| 6cq12 | 4505.6 | 0.08 | 0.00 | 0.00 | 0.00 | 0.00 | 0.00 | 2.49 | 0.08 |
| 6sr12 | 502.9 | 0.18 | 0.12 | 0.00 | 0.00 | 0.00 | 0.00 | 2.62 | 0.08 |
| 6cq15 | 5133.4 | 0.58 | 0.00 | 0.09 | 0.08 | 0.06 | 0.13 | 4.98 | 0.23 |
| 6sr15 | 1654.6 | 1.12 | 0.24 | 0.42 | 0.19 | 0.24 | 0.43 | 4.93 | 1.21 |
| 6cq18 | 5765.5 | 1.57 | 0.42 | 0.50 | 0.51 | 0.22 | 0.42 | 6.55 | 1.87 |
| 6sr18 | 1856.3 | 2.33 | 0.39 | 0.68 | 1.07 | 0.77 | 0.85 | 6.62 | 1.93 |
| All avg. | | 11.77 | 1.77 | 1.61 | 1.61 | 1.81 | 1.79 | 6.39 | 2.97 |
| Clique avg. | | 10.36 | 1.35 | 1.31 | 1.35 | 1.52 | 1.60 | 5.74 | 2.58 |
| SR avg. | | 13.19 | 2.20 | 1.91 | 1.88 | 2.10 | 1.97 | 7.03 | 3.37 |
| 3-AP avg. | | 33.84 | 4.14 | 4.20 | 4.20 | 4.51 | 4.56 | 9.93 | 5.76 |
| 4-AP avg. | | 10.12 | 2.39 | 1.68 | 1.66 | 2.22 | 1.89 | 6.59 | 3.95 |
| 5-AP avg. | | 2.16 | 0.36 | 0.28 | 0.29 | 0.30 | 0.40 | 4.32 | 1.27 |
| 6-AP avg. | | 0.98 | 0.19 | 0.28 | 0.31 | 0.21 | 0.30 | 4.70 | 0.90 |
| Small avg. | | 3.63 | 0.03 | 0.05 | 0.03 | 0.07 | 0.06 | 1.99 | 0.18 |
| Moderate avg. | | 14.18 | 1.97 | 1.59 | 1.56 | 1.89 | 1.70 | 7.10 | 2.71 |
| Large avg. | | 17.51 | 3.31 | 3.20 | 3.25 | 3.47 | 3.60 | 10.07 | 6.03 |

Table 2: Metaheuristics comparison.

| | Relative solution error, % | | | | | | | | |
|---|---|---|---|---|---|---|---|---|---|
| | 0.3 sec. | | | 1 sec. | | | 3 sec. | | |
| Inst. | HL | SA | GK | HL | SA | GK | HL | SA | GK |
| 3r40 | 49.75 | 120.00 | 10.75 | 44.25 | 99.00 | 9.75 | 41.50 | 84.50 | 6.25 |
| 3r70 | 512.86 | 102.86 | 3.29 | 18.14 | 82.86 | 1.71 | 16.86 | 72.71 | 0.71 |
| 3r100 | 5051.50 | 100.30 | 1.10 | 15.40 | 70.10 | 0.20 | 4.90 | 59.20 | 0.00 |
| 4r20 | 73.50 | 153.50 | 6.00 | 71.00 | 133.00 | 0.50 | 59.00 | 100.50 | 0.00 |
| 4r30 | 56.67 | 126.33 | 2.00 | 50.33 | 114.00 | 0.00 | 45.00 | 94.00 | 0.00 |
| 4r40 | 38.00 | 121.75 | 0.75 | 33.00 | 110.75 | 0.00 | 28.75 | 91.50 | 0.00 |
| 5r15 | 75.33 | 163.33 | 0.67 | 63.33 | 126.67 | 0.00 | 52.00 | 124.00 | 0.00 |
| 5r18 | 72.22 | 158.33 | 0.56 | 62.78 | 139.44 | 0.00 | 53.89 | 107.78 | 0.00 |
| 5r25 | 60.40 | 164.00 | 0.40 | 51.20 | 118.80 | 0.00 | 44.80 | 103.60 | 0.00 |
| 6r12 | 76.67 | 184.17 | 0.00 | 62.50 | 115.00 | 0.00 | 48.33 | 110.83 | 0.00 |
| 6r15 | 72.00 | 154.00 | 0.00 | 50.67 | 130.67 | 0.00 | 45.33 | 105.33 | 0.00 |
| 6r18 | 62.22 | 176.67 | 0.00 | 55.00 | 126.11 | 0.00 | 45.00 | 107.22 | 0.00 |
| All avg. | 516.76 | 143.77 | 2.13 | 48.13 | 113.87 | 1.01 | 40.45 | 96.77 | 0.58 |
| 3-AP avg. | 1871.37 | 107.72 | 5.05 | 25.93 | 83.99 | 3.89 | 21.09 | 72.14 | 2.32 |
| 4-AP avg. | 56.06 | 133.86 | 2.92 | 51.44 | 119.25 | 0.17 | 44.25 | 95.33 | 0.00 |
| 5-AP avg. | 69.32 | 161.89 | 0.54 | 59.10 | 128.30 | 0.00 | 50.23 | 111.79 | 0.00 |
| 6-AP avg. | 70.30 | 171.61 | 0.00 | 56.06 | 123.93 | 0.00 | 46.22 | 107.80 | 0.00 |
| Small avg. | 68.81 | 155.25 | 4.35 | 60.27 | 118.42 | 2.56 | 50.21 | 104.96 | 1.56 |
| Moderate avg. | 178.44 | 135.38 | 1.46 | 45.48 | 116.74 | 0.43 | 40.27 | 94.96 | 0.18 |
| Large avg. | 1303.03 | 140.68 | 0.56 | 38.65 | 106.44 | 0.05 | 30.86 | 90.38 | 0.00 |
| 3cq40 | 6.60 | 22.69 | 1.23 | 5.19 | 16.95 | 0.52 | 3.14 | 9.68 | 0.10 |
| 3sr40 | 6.55 | 27.10 | 1.87 | 5.11 | 18.18 | 0.74 | 4.44 | 15.92 | 0.07 |
| 3cq70 | 585.22 | 53.63 | 8.66 | 13.51 | 40.72 | 6.38 | 11.93 | 33.29 | 3.43 |
| 3sr70 | 744.70 | 58.53 | 8.97 | 15.63 | 44.69 | 7.15 | 15.00 | 39.52 | 5.70 |
| 3cq100 | 1013.95 | 68.28 | 11.94 | 1013.95 | 60.25 | 10.20 | 16.10 | 48.53 | 8.29 |
| 3sr100 | 1017.18 | 83.18 | 11.25 | 815.17 | 69.14 | 10.27 | 17.16 | 56.14 | 7.61 |
| 4cq20 | 1.71 | 15.53 | 0.07 | 1.35 | 12.28 | 0.03 | 0.87 | 10.48 | 0.03 |
| 4sr20 | 3.58 | 10.47 | 0.33 | 2.16 | 7.17 | 0.31 | 1.42 | 5.00 | 0.03 |
| 4cq30 | 7.51 | 30.65 | 2.66 | 6.66 | 21.57 | 0.91 | 5.64 | 18.21 | 0.69 |
| 4sr30 | 19.59 | 45.32 | 5.44 | 16.22 | 35.47 | 4.15 | 15.10 | 27.51 | 2.32 |
| 4cq40 | 17.90 | 37.87 | 6.80 | 11.60 | 34.76 | 4.46 | 10.41 | 28.53 | 2.47 |
| 4sr40 | 18.26 | 38.32 | 10.20 | 15.74 | 28.83 | 7.79 | 14.62 | 23.08 | 4.41 |
| 5cq15 | 0.95 | 30.11 | 0.07 | 0.41 | 29.80 | 0.03 | 0.20 | 28.66 | 0.00 |
| 5sr15 | 3.11 | 30.87 | 0.47 | 2.04 | 30.25 | 0.09 | 1.37 | 29.88 | 0.02 |
| 5cq18 | 2.41 | 38.73 | 0.57 | 2.17 | 38.26 | 0.20 | 1.27 | 36.40 | 0.04 |
| 5sr18 | 15.35 | 131.47 | 1.37 | 13.77 | 128.70 | 0.63 | 12.16 | 128.03 | 0.00 |
| 5cq25 | 7.52 | 48.11 | 3.84 | 6.11 | 45.41 | 1.97 | 5.00 | 45.06 | 0.54 |
| 5sr25 | 9.23 | 47.75 | 4.85 | 8.65 | 44.80 | 2.82 | 6.97 | 43.62 | 1.14 |
| 6cq12 | 0.62 | 35.66 | 0.24 | 0.08 | 35.55 | 0.00 | 0.01 | 35.18 | 0.00 |
| 6sr12 | 7.91 | 111.81 | 0.18 | 6.64 | 110.34 | 0.04 | 5.67 | 109.96 | 0.00 |
| 6cq15 | 2.26 | 43.66 | 1.43 | 1.58 | 43.68 | 0.32 | 1.31 | 42.22 | 0.08 |
| 6sr15 | 3.05 | 40.14 | 1.94 | 2.34 | 39.75 | 0.86 | 1.72 | 39.68 | 0.19 |
| 6cq18 | 3.91 | 51.19 | 15.43 | 2.48 | 49.98 | 1.43 | 1.90 | 48.95 | 0.51 |
| 6sr18 | 5.83 | 48.13 | 13.20 | 4.92 | 47.52 | 2.02 | 3.93 | 47.38 | 1.07 |
| All avg. | 146.04 | 47.88 | 4.71 | 82.23 | 43.09 | 2.64 | 6.56 | 39.62 | 1.61 |
| Clique avg. | 137.55 | 39.68 | 4.41 | 88.76 | 35.77 | 2.20 | 4.82 | 32.10 | 1.35 |
| SR avg. | 154.53 | 56.09 | 5.01 | 75.70 | 50.40 | 3.07 | 8.30 | 47.14 | 1.88 |
| 3-AP avg. | 562.37 | 52.24 | 7.32 | 311.43 | 41.66 | 5.88 | 11.30 | 33.85 | 4.20 |
| 4-AP avg. | 11.43 | 29.69 | 4.25 | 8.95 | 23.35 | 2.94 | 8.01 | 18.80 | 1.66 |
| 5-AP avg. | 6.43 | 54.51 | 1.86 | 5.52 | 52.87 | 0.96 | 4.49 | 51.94 | 0.29 |
| 6-AP avg. | 3.93 | 55.10 | 5.40 | 3.01 | 54.47 | 0.78 | 2.42 | 53.89 | 0.31 |
| Small avg. | 3.88 | 35.53 | 0.56 | 2.87 | 32.56 | 0.22 | 2.14 | 30.60 | 0.03 |
| Moderate avg. | 172.51 | 55.27 | 3.88 | 8.98 | 49.11 | 2.58 | 8.02 | 45.61 | 1.56 |
| Large avg. | 261.72 | 52.85 | 9.69 | 234.83 | 47.59 | 5.12 | 9.51 | 42.66 | 3.25 |

Table 3: Metaheuristics comparison.

| | Relative solution error, % | | | | | | | | | | | |
|---|---|---|---|---|---|---|---|---|---|---|---|---|
| | 10 sec. | | | 30 sec. | | | 100 sec. | | | 300 sec. | | |
| Inst. | HL | SA | GK | HL | SA | GK | HL | SA | GK | HL | SA | GK |
| 3r40 | 38.25 | 63.50 | 4.50 | 32.50 | 60.75 | 4.75 | 28.75 | 51.75 | 2.50 | 27.25 | 47.00 | 1.75 |
| 3r70 | 14.00 | 55.00 | 0.57 | 13.29 | 45.14 | 0.00 | 11.43 | 37.71 | 0.00 | 10.71 | 34.29 | 0.00 |
| 3r100 | 4.10 | 45.60 | 0.00 | 3.50 | 36.60 | 0.00 | 3.00 | 30.80 | 0.00 | 2.40 | 24.80 | 0.00 |
| 4r20 | 49.50 | 94.50 | 0.00 | 44.00 | 80.00 | 0.00 | 38.50 | 63.00 | 0.00 | 34.00 | 52.00 | 0.00 |
| 4r30 | 37.33 | 83.00 | 0.00 | 33.67 | 68.00 | 0.00 | 31.00 | 58.00 | 0.00 | 28.00 | 45.00 | 0.00 |
| 4r40 | 27.00 | 66.00 | 0.00 | 22.75 | 62.25 | 0.00 | 20.25 | 49.75 | 0.00 | 19.50 | 41.75 | 0.00 |
| 5r15 | 42.67 | 82.00 | 0.00 | 35.33 | 75.33 | 0.00 | 32.00 | 65.33 | 0.00 | 28.00 | 51.33 | 0.00 |
| 5r18 | 47.22 | 95.56 | 0.00 | 41.67 | 71.11 | 0.00 | 31.67 | 62.22 | 0.00 | 28.33 | 59.44 | 0.00 |
| 5r25 | 40.00 | 90.00 | 0.00 | 32.00 | 68.40 | 0.00 | 27.60 | 61.20 | 0.00 | 24.40 | 51.20 | 0.00 |
| 6r12 | 42.50 | 91.67 | 0.00 | 33.33 | 74.17 | 0.00 | 25.00 | 60.83 | 0.00 | 16.67 | 53.33 | 0.00 |
| 6r15 | 38.00 | 90.00 | 0.00 | 34.00 | 74.00 | 0.00 | 27.33 | 64.00 | 0.00 | 26.00 | 52.00 | 0.00 |
| 6r18 | 37.78 | 95.00 | 0.00 | 33.89 | 76.11 | 0.00 | 28.89 | 72.22 | 0.00 | 23.89 | 56.67 | 0.00 |
| All avg. | 34.86 | 79.32 | 0.42 | 29.99 | 65.99 | 0.40 | 25.45 | 56.40 | 0.21 | 22.43 | 47.40 | 0.15 |
| 3-AP avg. | 18.78 | 54.70 | 1.69 | 16.43 | 47.50 | 1.58 | 14.39 | 40.09 | 0.83 | 13.45 | 35.36 | 0.58 |
| 4-AP avg. | 37.94 | 81.17 | 0.00 | 33.47 | 70.08 | 0.00 | 29.92 | 56.92 | 0.00 | 27.17 | 46.25 | 0.00 |
| 5-AP avg. | 43.30 | 89.19 | 0.00 | 36.33 | 71.61 | 0.00 | 30.42 | 62.92 | 0.00 | 26.91 | 53.99 | 0.00 |
| 6-AP avg. | 39.43 | 92.22 | 0.00 | 33.74 | 74.76 | 0.00 | 27.07 | 65.69 | 0.00 | 22.19 | 54.00 | 0.00 |
| Small avg. | 43.23 | 82.92 | 1.13 | 36.29 | 72.56 | 1.19 | 31.06 | 60.23 | 0.63 | 26.48 | 50.92 | 0.44 |
| Moderate avg. | 34.14 | 80.89 | 0.14 | 30.65 | 64.56 | 0.00 | 25.36 | 55.48 | 0.00 | 23.26 | 47.68 | 0.00 |
| Large avg. | 27.22 | 74.15 | 0.00 | 23.03 | 60.84 | 0.00 | 19.93 | 53.49 | 0.00 | 17.55 | 43.60 | 0.00 |
| 3cq40 | 2.21 | 7.71 | 0.00 | 1.83 | 6.50 | 0.00 | 0.96 | 4.29 | 0.00 | 0.91 | 2.94 | 0.00 |
| 3sr40 | 3.16 | 9.61 | 0.11 | 2.41 | 6.63 | 0.00 | 1.82 | 5.27 | 0.00 | 1.00 | 3.10 | 0.00 |
| 3cq70 | 10.59 | 25.66 | 3.25 | 9.96 | 22.15 | 1.49 | 8.62 | 17.36 | 1.17 | 8.07 | 13.35 | 0.70 |
| 3sr70 | 12.87 | 32.42 | 3.11 | 12.17 | 24.37 | 1.86 | 10.50 | 20.82 | 1.18 | 9.89 | 17.23 | 0.41 |
| 3cq100 | 14.36 | 40.41 | 7.21 | 13.90 | 33.08 | 5.49 | 12.87 | 27.63 | 5.18 | 11.51 | 24.22 | 4.71 |
| 3sr100 | 15.34 | 45.72 | 6.27 | 14.03 | 38.22 | 4.63 | 13.22 | 30.96 | 3.30 | 12.56 | 27.57 | 3.27 |
| 4cq20 | 0.43 | 6.90 | 0.01 | 0.22 | 3.23 | 0.00 | 0.05 | 2.80 | 0.00 | 0.01 | 1.03 | 0.00 |
| 4sr20 | 0.91 | 2.04 | 0.03 | 0.69 | 1.54 | 0.00 | 0.48 | 1.19 | 0.00 | 0.22 | 0.40 | 0.00 |
| 4cq30 | 4.78 | 14.91 | 0.18 | 4.03 | 11.24 | 0.17 | 3.03 | 7.29 | 0.14 | 2.45 | 4.63 | 0.07 |
| 4sr30 | 13.40 | 22.09 | 1.01 | 12.24 | 18.87 | 0.52 | 10.60 | 14.32 | 0.28 | 9.68 | 11.19 | 0.13 |
| 4cq40 | 9.43 | 20.53 | 1.02 | 8.92 | 15.90 | 0.87 | 8.40 | 12.47 | 0.39 | 7.67 | 7.79 | 0.45 |
| 4sr40 | 13.43 | 17.58 | 1.85 | 11.84 | 14.93 | 1.30 | 10.48 | 12.03 | 0.47 | 10.14 | 9.56 | 0.41 |
| 5cq15 | 0.06 | 28.03 | 0.00 | 0.03 | 27.55 | 0.00 | 0.03 | 27.14 | 0.00 | 0.00 | 26.99 | 0.00 |
| 5sr15 | 0.56 | 29.75 | 0.00 | 0.19 | 29.66 | 0.00 | 0.00 | 29.66 | 0.00 | 0.00 | 29.66 | 0.00 |
| 5cq18 | 0.78 | 34.26 | 0.04 | 0.38 | 33.27 | 0.02 | 0.03 | 32.75 | 0.00 | 0.00 | 32.44 | 0.00 |
| 5sr18 | 10.24 | 126.20 | 0.06 | 8.81 | 124.74 | 0.00 | 7.09 | 123.65 | 0.00 | 6.28 | 123.09 | 0.00 |
| 5cq25 | 3.95 | 42.39 | 0.10 | 3.35 | 41.83 | 0.03 | 2.53 | 39.92 | 0.07 | 2.27 | 39.36 | 0.06 |
| 5sr25 | 6.21 | 43.15 | 0.64 | 5.85 | 42.13 | 0.31 | 5.16 | 41.85 | 0.10 | 4.37 | 41.70 | 0.14 |
| 6cq12 | 0.00 | 34.69 | 0.00 | 0.00 | 34.71 | 0.00 | 0.00 | 34.00 | 0.00 | 0.00 | 33.93 | 0.00 |
| 6sr12 | 4.16 | 109.43 | 0.00 | 3.66 | 109.41 | 0.00 | 2.11 | 109.41 | 0.00 | 1.59 | 109.41 | 0.00 |
| 6cq15 | 0.76 | 41.57 | 0.06 | 0.39 | 41.39 | 0.00 | 0.07 | 41.11 | 0.00 | 0.03 | 40.80 | 0.00 |
| 6sr15 | 1.29 | 39.57 | 0.22 | 0.93 | 39.57 | 0.09 | 0.70 | 39.57 | 0.00 | 0.60 | 39.57 | 0.00 |
| 6cq18 | 1.39 | 47.49 | 0.26 | 1.06 | 47.27 | 0.15 | 0.64 | 46.90 | 0.08 | 0.43 | 46.78 | 0.06 |
| 6sr18 | 3.31 | 47.23 | 0.25 | 2.79 | 47.15 | 0.03 | 2.14 | 47.14 | 0.04 | 1.74 | 47.14 | 0.04 |
| All avg. | 5.57 | 36.22 | 1.07 | 4.99 | 33.97 | 0.71 | 4.23 | 32.06 | 0.52 | 3.81 | 30.58 | 0.44 |
| Clique avg. | 4.06 | 28.72 | 1.01 | 3.67 | 26.51 | 0.69 | 3.10 | 24.47 | 0.59 | 2.78 | 22.85 | 0.50 |
| SR avg. | 7.07 | 43.73 | 1.13 | 6.30 | 41.43 | 0.73 | 5.36 | 39.66 | 0.45 | 4.84 | 38.30 | 0.37 |
| 3-AP avg. | 9.76 | 26.93 | 3.33 | 9.05 | 21.83 | 2.25 | 8.00 | 17.72 | 1.80 | 7.32 | 14.73 | 1.51 |
| 4-AP avg. | 7.06 | 14.01 | 0.68 | 6.32 | 10.95 | 0.48 | 5.51 | 8.35 | 0.21 | 5.03 | 5.77 | 0.18 |
| 5-AP avg. | 3.64 | 50.63 | 0.14 | 3.10 | 49.86 | 0.06 | 2.47 | 49.16 | 0.03 | 2.15 | 48.87 | 0.03 |
| 6-AP avg. | 1.82 | 53.33 | 0.13 | 1.47 | 53.25 | 0.05 | 0.94 | 53.02 | 0.02 | 0.73 | 52.94 | 0.02 |
| Small avg. | 1.44 | 28.52 | 0.02 | 1.13 | 27.40 | 0.00 | 0.68 | 26.72 | 0.00 | 0.47 | 25.93 | 0.00 |
| Moderate avg. | 6.84 | 42.09 | 0.99 | 6.11 | 39.45 | 0.52 | 5.08 | 37.11 | 0.35 | 4.62 | 35.29 | 0.16 |
| Large avg. | 8.43 | 38.06 | 2.20 | 7.72 | 35.06 | 1.60 | 6.93 | 32.36 | 1.20 | 6.34 | 30.51 | 1.14 |